\documentclass{article}
\usepackage[english]{babel}

\usepackage[letterpaper,top=2cm,bottom=2cm,left=3cm,right=3cm,marginparwidth=1.75cm]{geometry}
\usepackage{authblk}
\usepackage[numbers,comma,sort&compress]{natbib}
\usepackage{hypernat}
\usepackage{amsmath}
\usepackage{graphicx}
\usepackage[colorlinks=true, allcolors=blue]{hyperref}
\usepackage{placeins} 
\usepackage{tabularx}
\usepackage{tabu}
\usepackage{booktabs}
\usepackage{array}
\usepackage{makecell}
\usepackage{siunitx}
\usepackage{ulem}

\usepackage{float}
\usepackage{multicol}
\usepackage{dblfloatfix}
\usepackage{afterpage}
\usepackage{wrapfig}

\newcommand{\LSMO}{La$_{0.7}$Sr$_{0.3}$MnO$_3$}
\newcommand{\LCMO}{La$_{0.7}$Ca$_{0.3}$MnO$_3$}
\newcommand{\YBCO}{YBa$_2$Cu$_3$O$_7$}
\newcommand{\LSAT}{(LaAlO$_3$)$_{0.3}$(Sr$_2$TaAlO$_6$)$_{0.7}$}

\title{Long range proximity effects in planar structures involving the halfmetal ferromagnet \LSMO~and Pt interlayers.}
\author[1]{Junxiang Yao}
\affil[1]{Huygens-Kamerlingh Onnes Laboratory, Leiden Institute of Physics, \newline
Leiden University, P.O. Box 9504, 2300 RA Leiden, Netherlands}
\author[1]{Julian van Doorn}
\author[2]{Mariona Cabero}
\affil[2]{ICTS-Centro National de Microscopia Electronica, Universidad Complutense de Madrid, 28040 Madrid, Spain}
\author[1]{Jan Aarts\footnote{Author to whom correspondence should be addressed: aarts@physics.leidenuniv.nl}}

\date{\today}

\begin{document}

\maketitle

\begin{abstract}
Over the last decade, there has been steady research on superconducting junctions with a ferromagnet as the weak link, and where triplet correlations can transport supercurrents over a substantial distances. Of particular interest are halfmetallic ferromagnets, in which only one spin band is present, so that, presumably, the induced supercurrent is fully spin-polarized. We have earlier reported on a study of triplet transport in planar \LSMO (LSMO) nanostrip Josephson junctions with NbTi superconducting contacts, where we found high values for the supercurrents, and large junction lengths (up to 1.3~$\mu$m). Here, we extend that work by studying the dependence of the critical current I$_c$ on the length of the nanostrip between the contacts and the width of the strip. All junctions show strong supercurrents, but we do not observe simple systematics. Apparently, the fabrication process does not allow sufficient control over some of its parameters. To gain more insight in the mechanism for triplet generation at the LSMO/NbTi interface, we also studied the effect of Pt as an interlayer between the LSMO and the NbTi. For this, we etched a NbTi/Pt electrode structure on a full film of LSMO. The results are highly promising, showing sharp superconducting transitions and zero-resistance states being reached at an electrode distance of 2~$\mu$m, with indications that larger distances should be feasible.
\end{abstract}

\begin{multicols}{2}

\section{Introduction}
In order for Josephson junctions with a halfmetallic ferromagnetic (HMF) weak link to carry a supercurrent, the conventional spinless Cooper pairs in the superconducting (S) banks need to be converted to equal-spin triplets. These triplet pairs are able to carry supercurrents over micrometer distances, giving rise to the so-called long range proximity (LRP) effect. It stems from the fully spin-polarized nature of the HFM, in which the missing spin band makes triplet pair breaking  difficult. The singlet-triplet conversion requires some form of spin mixing and scattering at the S/HMF interface, which can be achieved by inserting additional ferromagnetic (F) layers, or simply by the presence of magnetic disorder. Long range effects in planar structures with HMF’s were reported in devices based on CrO$_2$, on \LSMO~(LSMO). and on \LCMO~(LCMO). With CrO$_2$, either a superconducting contact structure was put on top of a full CrO$_2$ film \cite{keizer2006spin,anwar2010long,anwar2011}, or stacked S/F/N contacts (N a normal metal) were fabricated on nanostrips \cite{singh2016high}. With LSMO, contacts of superconducting \YBCO~(YBCO), with a spacing of about 1~$\mu$m, were deposited by lithographic methods \cite{sanchez2022}. A similar method was used to study arrays of YBCO islands on an LCMO film \cite{sanchez2024}. \\
\begin{figure*}[t]
\centering
\includegraphics[width=.7\textwidth]{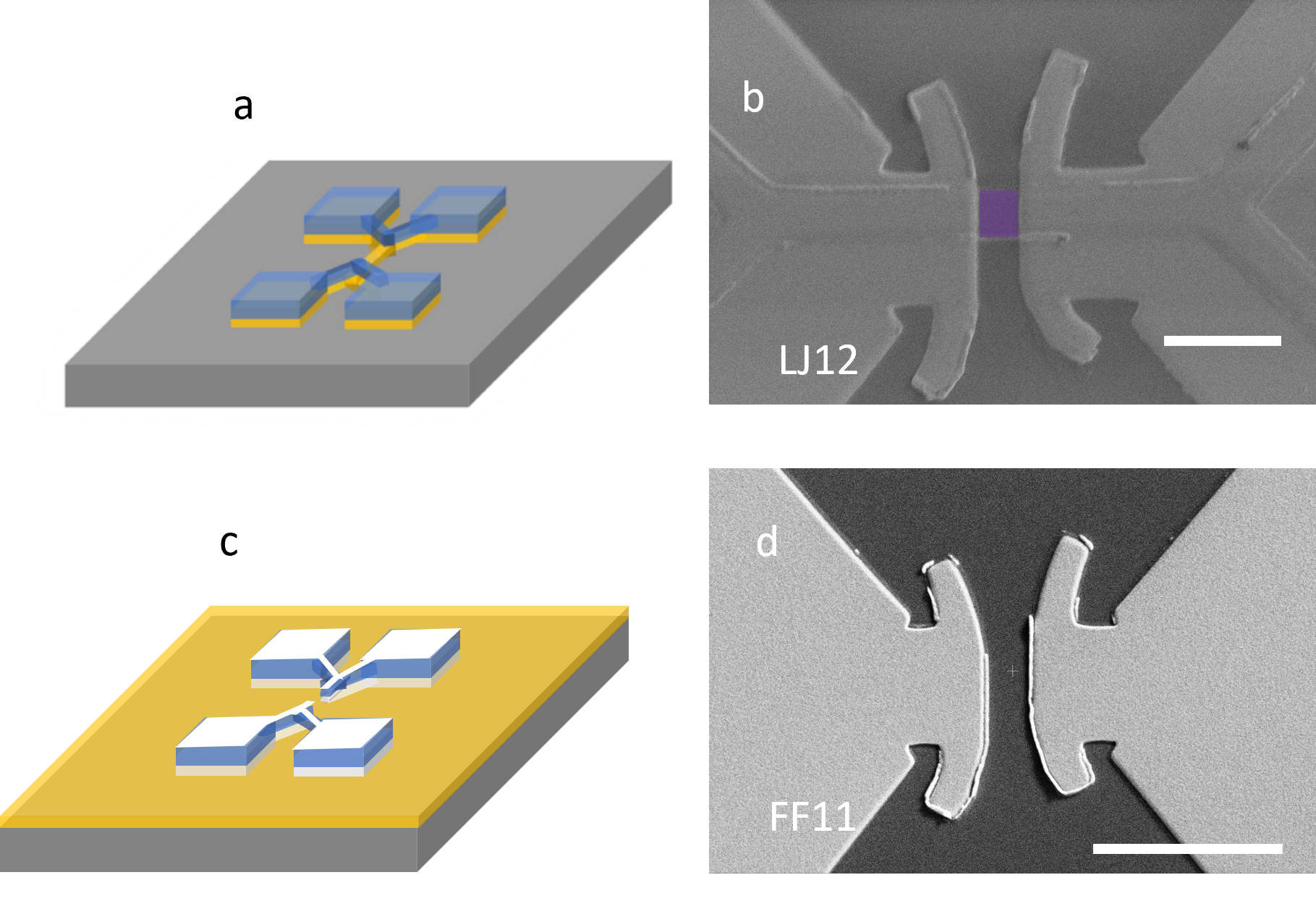}
\caption{Layout and SEM image of the LJ (upper row) and FF (lower row) devices. (a) NbTi top layer contacts (blue) on LSMO/LSAT, with an LSMO connecting bridge (yellow). (b) SEM picture of device LJ12 (L = 1.58~$\mu$m), with the LSMO bridge in false-color purple. The scale bar is 5~$\mu$m. (c) Top contacts on a NbTi (blue)/ Pt (white) double layer on a full film of LSMO (yellow) on LSAT. (d) SEM picture of device FF11 (L = 2.1~$\mu$m). The scale bar is 10~$\mu$m.}
\label{Fig1-DevFab}
\end{figure*}
Also with LSMO as the HFM, we recently reported LRP effects in junctions based on the conventional superconductor NbTi. The fabrication involved ion-etching of an LSMO/NbTi bilayer, using a Pt hard mask as described in Ref.~\cite{yao2024apl}. In particular, we found a large critical current $J_c$  of 1.1~$\times 10^9$ [A/m$^2$] in a 5~$\mu$m wide nanostrip with a length between the contacts $L$ = 1.3~$\mu$m, at 2~K, well below the junction critical temperature $T_c$ of 6.5~K. We noted at the time that even without building in a special spin scattering layer, the LSMO/NbTi interface is a strong triplet generator, as we also found in experiments on LSMO/NbTi disk- and square-shaped junctions with a small junction length \cite{yao2024prr}. Here, we extend our research in two directions. First, we fabricated a new series of long lateral junctions of LSMO/NbTi (called LJ devices), to find the dependence of $J_c$ on $L$. We do find strong critical currents, as in the earlier work, but not a simple monotonous dependence on $L$. This indicates a lack of control in the fabrication process, in particular when it comes to the LSMO/NbTi interface where the triplet generator quite likely resides.
\\
\noindent
To explore this further, we decided to change the character of the interface by separating the LSMO and the NbTi with a non-reactive normal metal. For that layer we chose Pt, since earlier experiments with Ag had shown the Ag to form islands \cite{yao2024prr}. The devices we made consisted of trilayers LSMO/Pt(7)/NbTi, with the Pt layer 7~nm thick. Instead of LJ devices with well defined LSMO bridges, we fabricated NbTi/Pt structures on top of full films of LSMO (called FF devices). They are similar to the experiments mentioned above on CrO$_2$ \cite{keizer2006spin,anwar2010long,anwar2011}, and fully adequate to demonstrate long range supercurrents, if they exist. The largest distance between the electrodes was 2.1~$\mu$m. All FF devices demonstrate clear proximity effects, meaning a zero resistance state, strong supercurrents, and a strong response to an out-of-plane (OOP) magnetic field. Triplet pairs must be involved, and apparently, the LSMO/NbTi interface is not a crucial ingredient in generating them. Below, we give the details of the experiments and discuss the results. \\

\noindent
\begin{table*}
\centering
    \begin{tabular}{|c|c|c||c|c|c||c|c|} \hline
2~$\mu$m    &      &            & 5~$\mu$m  &      &  &  &  \\ \hline
    LJ       & $L$    & $J_c$(2~K) & LJ       & $L$    & $J_c$(2~K) & FF & L \\
             & $\mu$m  & 10$^9$ [A/m$^2$]   &   & $\mu$m  & 10$^9$ [A/m$^2$] & & $\mu$m \\\hline
    8      & 0.47  &  2         & 5        & 0.76  &  2 & 7 & 0.9 \\
    3*      & 0.59  &  0.5       & 11       & 1.19 &  -- & 9 & 1.7 \\
    7      & 1.15 & 0.5        & 4*        & 1.35 &  1.1 & 11 & 2.1 \\
    12     & 1.58 & 0.9        &  --         &  --    &  --    &  --  &  --   \\
    \hline
    \end{tabular}
\caption{Basic parameters for the two types of LJ devices, with a bridge width of 2~$\mu$m and 5~$\mu$m, arranged by increasing length $L$ between the contacts. Given are the device number, the length $L$, and the critical current density $J_c$, measured at 2~K. The starred devices were reported on before \cite{yao2024apl}. Also given are the device numbers and length $L$ for the three FF devices.}
\label{table-devices}
\end{table*}

\section{Experimental}

The nanostrip devices and the full-film devices were fabricated from a bilayer and a trilayer, respectively, of NbTi/LSMO and Nb/Pt/LSMO. First, a 40~nm LSMO film was grown on a (001)-oriented \LSAT~(LSAT) single crystal substrate at 700 $^0$C in an off-axis high-pressure sputtering system in a flowing mixture of argon and oxygen gas (3:2), at a pressure of 0.7~mbar. Details of the procedure, and a characterization of the films grown in this way have been given in Refs.~\cite{yao2024prr,yao2024apl}. The NbTi and Pt/NbTi films were deposited in-situ after cooling the system to ambient, followed by pumping down to 10$^{-7}$~mbar. Sputtering was done in argon, at a pressure of 0.1~mbar. We chose a thickness of 60~nm for the NbTi film in the case of LSMO/NbTi. In the case of NbTi/Pt, the NbTi film was 70~nm, and the Pt was 7~nm. \\
The nanostrip devices were fabricated using the Pt hard mask technique described in Ref.~\cite{yao2024apl}. We made strips with widths of 2~$\mu$m and 5~$\mu$m, with values of $L$ ranging between 0.5~$\mu$m and 1.6~$\mu$m. The exact dimensions are given in Table~\ref{table-devices}. The full-film devices were fabricated as shown in Fig.~\ref{Fig1-DevFab}, by Ar-etching a contact structure on top of the 40~nm LSMO films. Fig.~\ref{Fig1-DevFab} also shows SEM pictures for the two types of devices, to underscore the difference. For the etching, we first determined the etch rates for NbTi (8~nm/min) and Pt (24~nm/min). Etching was performed for 9~minutes. Three devices were measured, called FF7, FF9, and FF11, with contact distances of 0.9~$\mu$m, 1.7~$\mu$m, and 2.1~$\mu$m After the first set of measurements, they were etched again for 2 minutes. For these devices, it is important that the Pt film is smooth, flat, and fully wetting the LSMO. This was investigated by Atomic Force Microscopy (AFM). The morphology of the LSMO/Pt interface is also important, since magnetic disorder may play a role in generating triplets. We used Aberration corrected scanning transmission electron microscopy (STEM) combined with electron energy loss spectroscopy (EELS) to study the structural and chemical properties of the sample. The STEM-EELS data were acquired at 200 kV in a JEOL ARM200cF microscope equipped with a spherical aberration corrector and a Gatan Quantum Dual-EELS spectrometer. \\
Transport measurements were performed with a Physical Properties Measurement Platform (Quantum Design). We measured resistance and (critical) current in zero applied magnetic field, as well as so-called superconducting quantum interference patterns (SQI's). We define the $z$-direction as out-of-plane (OOP), the $x$-direction as in-plane (IP) along the axis connecting the contacts, and the $y$-direction as IP and perpendicular to the axis connecting the contacts. \\

\section{Results}
\subsection{The long junction devices}
\begin{figure*}[t]
\centering
\includegraphics[width=\textwidth]{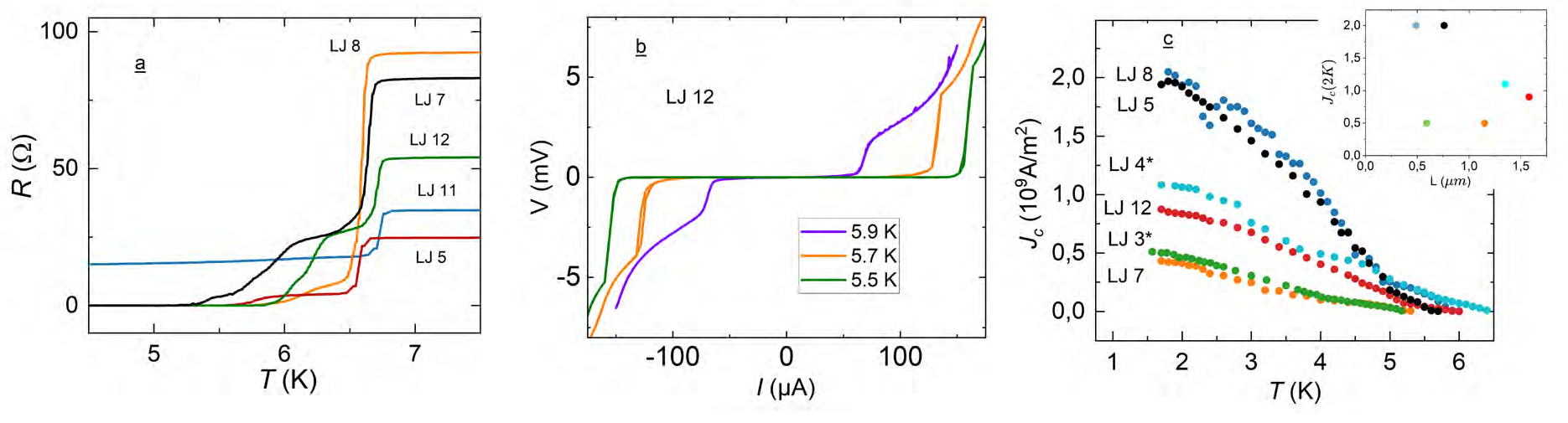}
\caption{Transport properties of the LJ devices. (a) Resistance $R$ versus temperature $T$ for the devices as indicated. (b) Current $I$ versus voltage $V$ for LJ12 ($L$ = 1.58~$\mu$m) at 5.9~K, 5.7~K and 5.5~K. (c) Critical current $J_c$ as function of temperature $T$ for the LJ devices as indicated, with the highest temperature showing the lowest $J_c$. The starred devices were reported on before \cite{yao2024apl}. The inset shows the behavior of $J_c$ at 2~K as funciont of $L$, color-coded in the same manner as the main plot. }
\label{Fig2-RT-jc}
\end{figure*}

Fig.~\ref{Fig2-RT-jc}a shows resistance $R$ versus temperature $T$ for the newly fabricated devices. We observe the well-known signature of a proximized structure, where the NbTi banks become superconducting first (around 6.7~K), followed by a plateau that is determined by the resistance of the weak link, and then a resistance drop to zero when the link becomes fully superconducting. The relatively low T$_c$ of the NbTi is due to a background pressure of 10$^{-7}$~mbar and the fact that NbTi is a good getter. The smaller-length devices LJ5 and LJ8 ($L$ $\approx$ 0.5~$\mu$m) show a lower resistance at the plateau, even though their width is different. The longer devices LJ7, LJ11 and LJ12 show a higher plateau resistance, but LJ11 does not become superconducting. Critical currents were determined from current-voltage ($I$-$V$) characteristics, as shown for the longest device, LJ12, in Fig~\ref{Fig2-RT-jc}b. The curves are somewhat rounded, which we know is due to imperfect filtering of the signal lines, but a critical current can be readily defined.
We used an area $A$ of 40~nm $\times w$ ($w$ being the width, either 2~$\mu$m or 5~$\mu$m) to compute critical current densities from the measured currents. Figure~\ref{Fig2-RT-jc}c shows the temperature dependence of the critical current densities $J_c$ for both the new and the old devices (designated with a *). The critical currents at 2~K are very high. For LJ12 ($L$=1.6~$\mu$m), $J_c$ is of the order of 10$^9$ [A/m$^2$]. However, we also note a stark lack of systematic behavior. In the new batch, $J_c$ of LJ7 ($L$=1.1~$\mu$m) is two times lower than $J_c$ of LJ12 ($L$=1.6~$\mu$m). Comparing with the older batch, $J_c$ of LJ8 ($L$=0.47~$\mu$m) is four times larger than $J_c$ of the similar-length LJ3* ($L$=0.59~$\mu$m), and roughly equal to $J_c$ of LJ5, which is 2.5 times wider. The experiments confirm the earlier ones in a qualitative way, and show undiminished promise for transporting spin-polarized supercurrents over long lengths, but they also signal a lack of control in the fabrication.\\
We also measured SQI's. Figure~\ref{Fig4-SQIs}a shows a measurement on LJ12, with the field applied out-of-plane (OOP) with respect to the sample surface (the $z$-direction). The data were taken at 5.65~K, quite close to T$_c$. They show the characteristic peak shape we found before, to be discussed below.

\subsection{Full film devices}
\begin{figure*}
\centering
\includegraphics[width=\textwidth]{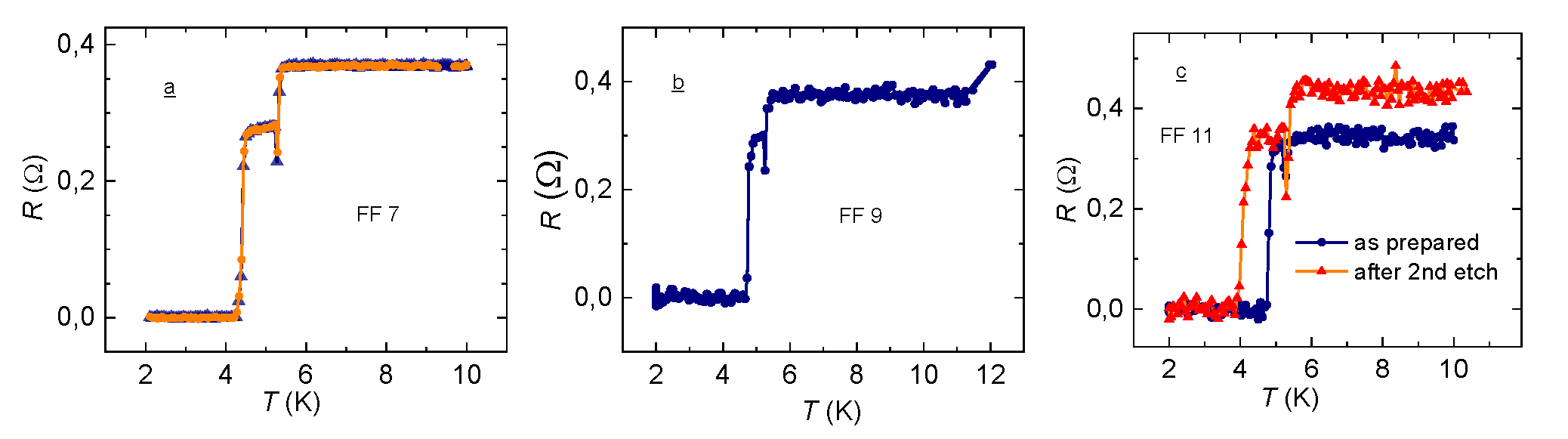}
\caption{Resistance $R$ versus temperature $T$ for the three full film devices (a) FF7, (b) FF9 and (c) FF11, discussed in the text. In (a) two traces are shown, one for increasing temperature (orange circles), and one for decreasing temperature (blue triangles). In (c) the change in R(T) after another 2 minute Ar-etch step is also shown. The measurements were performed with an applied current of 1~$\mu$A.}
\label{Fig3-RT-FF}
\end{figure*}

\begin{figure*}[b]
\centering
\includegraphics[width=\textwidth]{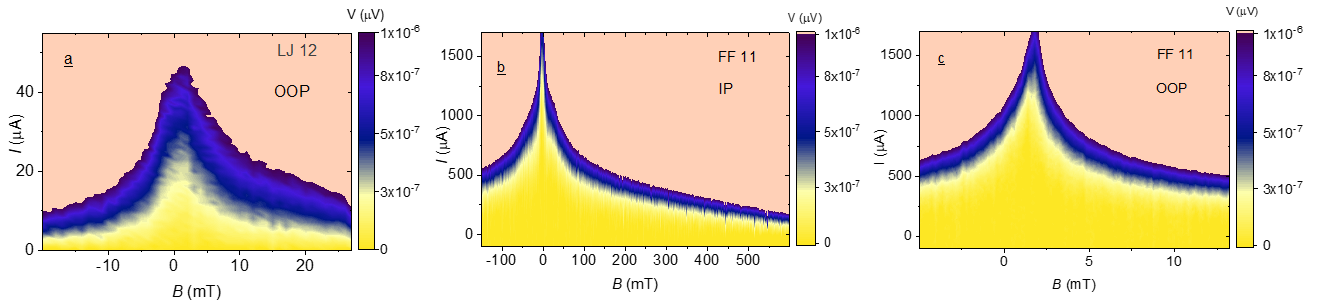}
\caption{Superconducting Quantum Interference (SQI) patterns, represented as a color plot of voltage $V$ versus applied current $I$, for (a) LJ12, measured with an OOP field ($z$-direction) at 5.65~K, (b) FF11 with an IP field along $y$ (perpendicular to the axis between the contacts), at 4.65~K, and (c) FF11 with an OOP field, at 4.65~K.}
\label{Fig4-SQIs}
\end{figure*}

Since we believe the triplet generator to reside at the interface, we next prepared a set of devices where a thin Pt layer is inserted between the LSMO and the NbTi layers. Rather than attempting to fabricate long junctions, we prepared full film devices as described above, since they can quickly answer the question whether supercurrents are generated. Measurements of $R(T)$ on devices FF7 ($L$=0.9~$\mu$m), FF9 ($L$=1.7~$\mu$m) and FF11 ($L$=2.1~$\mu$m) are given in Fig.~\ref{Fig3-RT-FF}. They all show the same normal state resistance $R_N$ of about 0.4~$\Omega$, and a sharp drop around 6~K, where presumably the contacts become superconducting. The resistance does not drop to 0, but shows a sharp dip, coming back to a plateau at a value which is about 0.1~$\Omega$ lower than $R_N$. The plateau has a width of typically 1~K, after which, between 4~K and 5~K, the resistance does drop down to 0. We do not fully understand the behavior, but the plateau as main ingredient is similar to what is generally observed in lateral proximity structures. The details of the behavior must be due to the multiple current paths, furnished by the full films. For FF11, we also show the effect of an extra Ar-etch of 2 minutes. $R_N$ has clearly increased, which we tend to ascribe to the film having become thinner, and the plateau width has increased such that zero resistance is now reached just below 4~K. \\
SQI patterns for FF11, taken at 4.65~K, are shown in Fig.~\ref{Fig4-SQIs}b,c for the OOP configuration, and with the field in-plane and perpendicular to the axis connecting the contacts ($y$-direction). The currents are very high, and only measurable close to T$_c$. There is a clear difference in field sensitivity, and again, the SQI's are sharply peaked. \\
We note in passing that this non-Gaussian lineshape is uncommon. For diffusive long and narrow SNS junctions, the SQI is typically Gaussian, as discussed for instance in Ref.~\cite{chiodi12}. We find strongly peaked patterns both for the LJ and FF devices, indicating that the dephasing over different trajectories by the flux is not fully random. A complicating factor is that also the magnetization of the film or strip is changing when applying the magnetic field. Understanding this behavior will require a dedicated study.
\begin{figure*}[t]
\centering
\includegraphics[width=0.7 \textwidth]{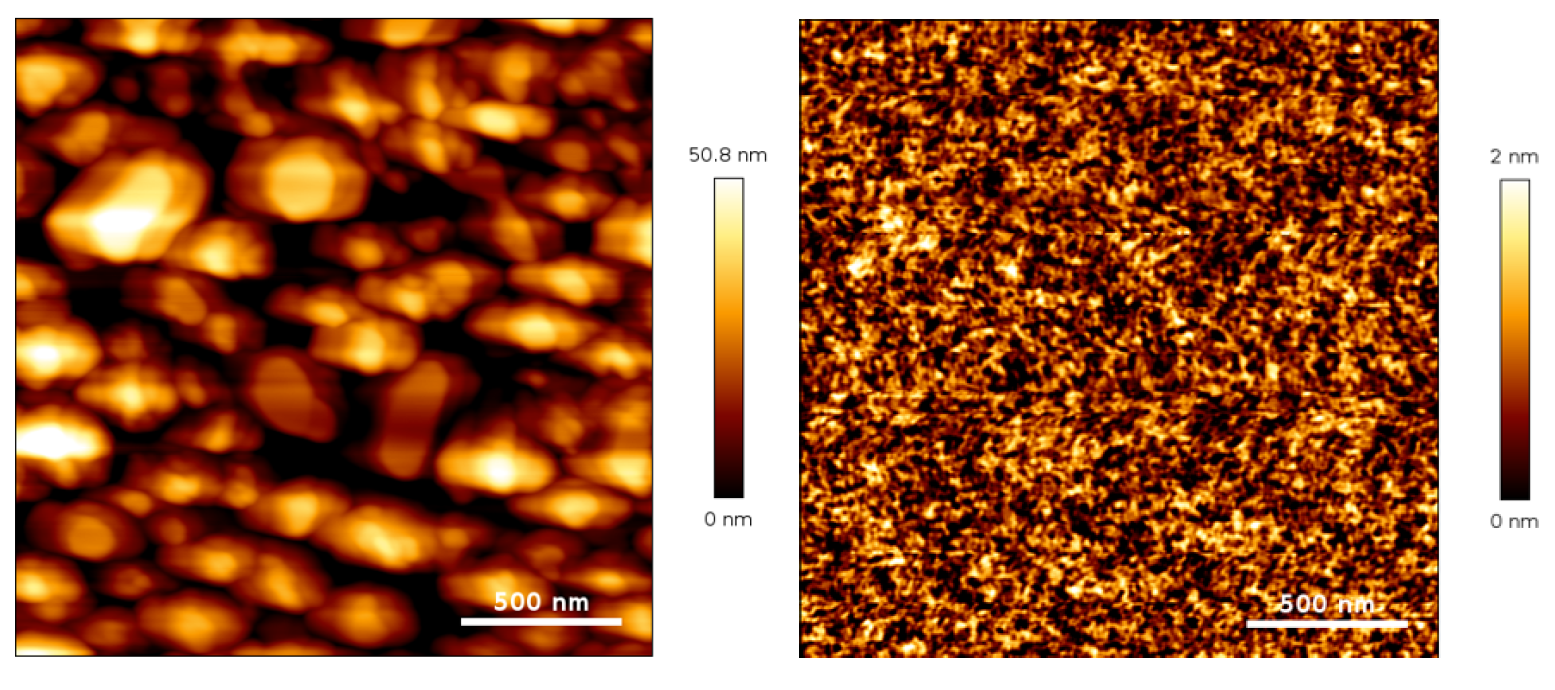}
\caption{AFM images of the surface of (a) an LSMO/Ag bilayer and (b) an LSMO/Pt bilayer. In the case of Ag, high roughness and island formation is visible. In the case of Pt, the roughness is no more than 2~nm.}
\label{Fig5-AFM}
\end{figure*}
\subsection{Growing Pt on LSMO}
\begin{figure*}[t]
\centering
\includegraphics[width= 0.6 \textwidth]{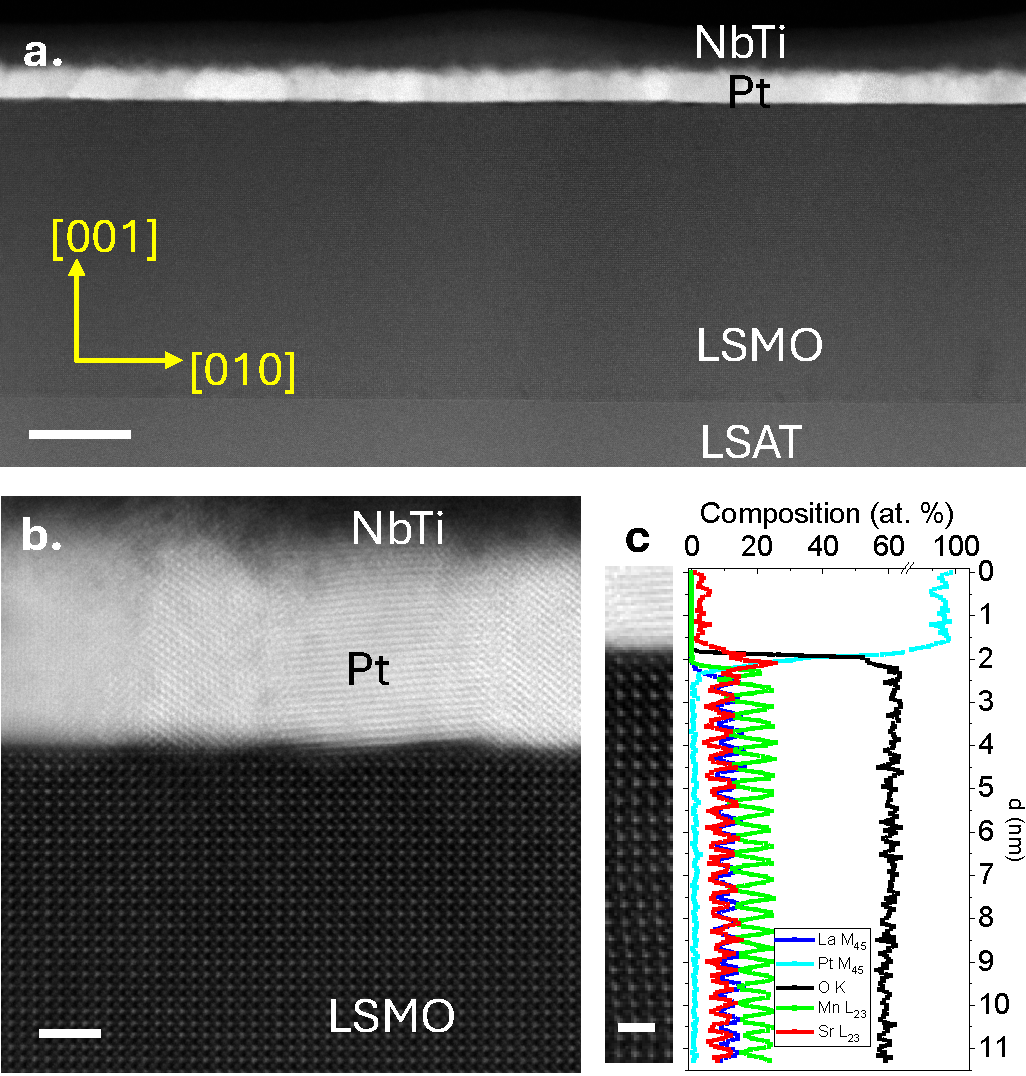}
\caption{STEM EELS characterization of an LSMO/Pt/NbTi sample grown on an LSAT substrate. (a) Low magnification high angle annular dark field (HAADF) image showing the sharp and coherent heterostructure. (b) High resolution HAADF image of the atomically sharp LSMO/Pt interface. (c) Left: annular dark field (ADF) image acquired simultaneously with an EELS spectrum image on the same region of the sample. Right: relative atomic percentage profiles obtained by lateral averaging over the same spectrum image, using the Sr L2,3, Mn L2,3, La M4,5, O K, and Pt M4,5 edges. Scale bars: 20 nm, 2 nm, and 1 nm in (a), (b), and (c), respectively.
}
\label{Fig6-TEM}
\end{figure*}

In earlier work we reported on attempting to use Ag as the interlayer, but found that Ag wets insufficiently, which leads to island formation and a layer which is not closed \cite{yao2024prr}. For Pt, this turns out to be very different. Figure~\ref{Fig5-AFM} shows AFM measurements on the surface of an LSMO/Ag and an LSMO/Pt bilayer. In the LSMO/Ag image, islands are clearly visible, and the roughness is of order 50~nm. In the LSMO/Pt image, the surface is quite flat and homogeneous, with a roughness of no more than 2~nm. This is fully corroborated by transmission electron microscopy experiments.
The internal structure of the LSMO/Pt/NbTi heterostructure was examined by aberration corrected STEM-EELS, as shown in Fig.~\ref{Fig6-TEM}. In the cross section high-angular annular dark-field (HAADF image) of Fig.~\ref{Fig6-TEM}a, the multilayer sequence is clearly resolved: the LSMO film remains epitaxial on the LSAT substrate, and the metallic Pt and NbTi overlayers display a laterally uniform thickness across the field of view. At higher magnification, the LSMO/Pt boundary in Fig.~\ref{Fig6-TEM}b appears as an abrupt transition from the ordered perovskite lattice to the polycrystalline Pt layer, with no discernible interfacial roughening or secondary phases at the nanometer scale. To gain chemical insight at the interface, an EELS spectrum image was acquired across the LSMO/Pt region, as illustrated in Fig.~\ref{Fig6-TEM}c. The annular dark field image on the left serves as a reference for the area analyzed, while the relative atomic percentage quantification on the right is obtained by laterally averaging the spectrum image and using the Sr L2,3, Mn L2,3, La M4,5, O K, and Pt M4,5 edges. This quantification confirms the expected atomic step like increase of Pt at the metal layer and the corresponding decrease of the oxide cations, demonstrating that the Pt film fully covers the LSMO. On the LSMO side, the interfacial composition is enriched in Sr and depleted in La, while deeper into the film the La/Sr ratio and Mn content approach their nominal values, consistent with a perovskite structure terminating in an Sr rich layer. The oxygen quantification follows the oxide stoichiometry up to the interface, without indications of a broad oxygen deficient region, supporting the picture of a chemically sharp and well preserved LSMO/Pt junction. Taken together, the interface shows none of the atomic mixing that we observed in the case of LSMO/NbTi. It does not point to the presence of magnetic disorder, although it is possible that the last Mn layer contains Mn tending more to 4+, given the absence of La in the terminating layer.
\section{Discussion and Conclusions}
The data we presented on the long junctions allow to draw several conclusions. One is that the LSMO/NbTi stack is able to generate triplets, leading to quite strong supercurrents in devices made at different times. Also, we confirmed that the supercurrent densities are high: for $T_c$'s around 6~K, $J_c$(2K) is of order 10$^9$~A/m$^2$, even for a distance of 2~$\mu$m. On the other hand, triplet generation is not reproducible enough to yield a reliable dependence of $J_c$ on $L$. From what we know about the disorder that exists at the LSMO/NbTi interface, this may not be surprising. If the interface itself is the triplet generator, for instance due to Mn valence variations and ensuing magnetic disorder, it is not a well-controlled one. If the generator lies deeper in the LSMO layer, for instance by a slow variation of the magnetic order, the disordered interface is a badly controlled feeder of singlets to that hypothetical spin-mixing layer. \\ \\
The experiments on the full film devices were meant to shed more light on this question. By removing the direct contact between NbTi and LSMO, and preventing the formation of the magnetically disordered interface, the triplet generator might actually have been removed. Instead, we still find a triplet generation mechanism, apparently promoted by the Pt layer. One possible, and even plausible, explanation comes from considering the effects of spin-orbit coupling (SOC). Over the years, this has been addressed theoretically (relevant treatments can be found in Refs.~\cite{bergeret13,bergeret14,jacobsen16,eskilt19}) and experimentally (without being exhaustive, see Refs.~\cite{banerjee18,bregazzi2024}). In particular, Refs.~\cite{bergeret14,eskilt19} treat our experimental layout, that of a lateral junction with a heavy metal interlayer between the superconducting contact and the (lateral) ferromagnetic weak link. The conclusion they reach is that, in the case of Rashba-like SOC and an in-plane magnetization, long range (LR) triplets are generated at the Pt/F interface and propagate in the $xy$-plane, while perpendicular to the interface (the $z$-direction) the triplets are short range. Experimentally, there are a number of different studies on the effect of incorporating Pt in S/F structures, but the picture is not yet fully clear. We will refrain here from reviewing other data, except for noting one experiment on perpendicular junctions by Satchell and Birge \cite{satchell18}. They prepared vertical Josephson junctions with superconducting Nb contacts around a central Co-Ru-Co block, and inserted thin Pt layers between contacts and block. In such junctions they found no signs of LR triplets, in line with the prediction that such triplets are not generated in the direction perpendicular to the interface (in this case the Pt/Co interface). It is of interest to note that, if Rashba-SOC plays a role in the triplet generation, this may also play a part in cases where Pt is absent, in particular in the cases of LSMO/NbTi, and LSMO/YBCO. It is, after all, the symmetry breaking and the electrical potential gradient that cause the Rashba effect, irrespective of (although possibly aided by) the strength of the SOC in the bulk. We conclude that a Pt interlayer is very effective in interfacing oxide metals with conventional metals. In an S/Pt/HMF geometry, the Pt interlayer shows clear promise for the controlled fabrication of lateral Josephson junctions carrying a fully spin-polarized supercurrent. Further experiments should divulge whether the triplet generation mechanism in this case is based on Rashba spin-orbit coupling at the Pt/LSMO interface, as we currently believe. Such experiments would require careful consideration of the material science involved: different interlayers should be wetting, but not oxidizing, and allow variations of the spin mixing conductance that is supposedly a measure for the effectiveness of spin triplet generation.\\
\vfill\null
\columnbreak
\section*{Acknowledgements}
The authors thank Kaveh Lahabi for discussions, and Floriana Lombardi who reminded us of the wetting property of Pt on metallic oxides. Marcel Hesselberth, Luc Wigbout and Sander van Leeuwen are acknowledged for their support in device preparation. J.Y. was partially funded by China Scholarship Council (Grant No. 201808440424). The work is partly financed by the Dutch Research Council (NWO) through Projectruimte Grant No. 680.91.128 and Project No. OCENW. XS22.2.032. The work was further supported by EU Cost Action No. CA21144 (SUPERQUMAP). M.C. received support from the Spanish state research agency AEI through Grant No. PID2020-118078RB-I00 and from the Regional Government of Madrid CAM through SINERGICO Project No. Y2020/NMT-6661 CAIRO-CM. Electron microscopy observations were carried out at ICTSCNME in UCM. The authors acknowledge the ICTS-CNME for offering access to their instruments.
\newpage
\bibliographystyle{unsrtnat}
\bibliography{LRP2}

@article{sanchez2022,
  title={Extremely long-range, high-temperature {J}osephson coupling across a half-metallic ferromagnet},
  author={Sanchez-Manzano, D. and Mesoraca, S. and Cuellar, F. A. and Cabero, M. and Rouco, V. and Orfila, G. and Palermo, X. and Balan, A. and Marcano, L. and Sander, A. and others},
  journal={Nature Materials},
  volume={21},
  number={2},
  pages={188--194},
  year={2022},
  publisher={Nature Publishing Group}
}

@article{sanchez2024,
  title={Long-range superconducting proximity effect in {Y}{B}a$_2${C}u$_3${O}$_7$/{L}a$_{0.7}${C}a$_{0.3}${M}n{O}$_3$ weak link arrays},
  author={Sanchez-Manzano, D. and Mesoraca, S. and Rodriguez-Corvillo, S. and Lagarrigue, A. and Gallego, F. and Cuellar, F. A. and Sander, A. and Rivera-Calzada, A. and Valencia, S. and Villegas, E. and Leon, C. and Santamaria, J.},
  journal={Applied Physics Letters},
  volume={124},
  pages={222603},
  year={2024},
  publisher={AIP}
}

@article{yao2024prr,
  title={Fabrication of planar halfmetallic ferromagnetic {J}osephson junctions with long range coupling},
  author={Junxiang, Y. and Fermin, R. and Cabero, M. and Lahabi, K. and Aarts, J.},
  journal={Physical Review Research},
  volume={6},
  pages={043114},
  year={2024},
  doi={10.1103/PhysRevResearch.6.043114},
  publisher={APS}
}

@article{yao2024apl,
  title={Triplet supercurrents in lateral {J}osephson junctions with a half-metallic ferromagnet},
  author={Junxiang, Y. and Aarts, J.},
  journal={Applied Physics Letters},
  volume={124},
  pages={202601},
  year={2024},
  doi={10.1063/5.0210842},
  publisher={AIP}
}

@article{anwar2011,
  title={Inducing supercurrents in thin films of ferromagnetic {C}r{O}$_2$},
  author={Anwar, M. S. and Aarts, J.},
  journal={Superconductor Science and Technology},
  volume={24},
  number={2},
  pages={024016},
  year={2011},
  publisher={IOP Publishing}
}

@article{singh2016high,
  title={High-quality {C}r{O}$_2$ nanowires for dissipation-less spintronics},
  author={Singh, A. and Jansen, C. and Lahabi, K. and Aarts, J.},
  journal={Physical Review X},
  volume={6},
  number={4},
  pages={041012},
  year={2016},
  publisher={APS}
}

@article{keizer2006spin,
  title={A spin triplet supercurrent through the half-metallic ferromagnet {C}r{O}$_2$},
  author={Keizer, R. S and G{\"o}nnenwein, S.  T. B. and Klapwijk, T. M. and Miao, G. and Xiao, G. and Gupta, A.},
  journal={Nature},
  volume={439},
  number={7078},
  pages={825--827},
  year={2006},
  publisher={Nature Publishing Group UK London}
}

@article{anwar2010long,
  title={Long-range supercurrents through half-metallic ferromagnetic {C}r{O}$_2$},
  author={Anwar, M. S. and Czeschka, F. and Hesselberth, M. and Porcu, M. and Aarts, J.},
  journal={Physical Review B},
  volume={82},
  number={10},
  pages={100501},
  year={2010},
  publisher={APS}
}

@Article{bergeret13,
  author        = {Bergeret, F. S. and Tokatly, I. V.},
  title         = {Singlet-Triplet Conversion and the Long-Range Proximity Effect in Superconductor-Ferromagnet Structures with Generic Spin Dependent Fields},
  journal       = {Physical Review Letters},
  year          = {2013},
  volume        = {110},
  pages         = {117003},
  publisher     = {American Physical Society},
  doi         = {10.1103/PhysRevLett.110.117003}
}

@article{bergeret14,
  author={Bergeret, F. S. and Tokatly, I. V.},
  title={Spin-orbit coupling as a source of long-range triplet proximity effect in superconductor-ferromagnet hybrid structures},
  journal={Physical Review B},
  volume={89},
  pages={134517},
  year={2014},
  publisher={APS},
  doi = {10.1103/PhysRevB.89.134517}
}

@article{jacobsen16,
  author={Jacobsen, S. H. and Kulagina, I. and Linder, J.},
  title={Controlling superconducting spin flow with spin-flip immunity using a
single homogeneous ferromagnet},
  journal={Scientific Reports},
  volume={6},
  pages={23926},
  year={2016},
  publisher={Nature},
  doi = {10.1038/srep23926 (2016).}
}

@article{eskilt19,
  author={Eskilt, J. R. and Amundsen, M. and Banerjee, N. and Linder, J.},
  title={Long-ranged triplet supercurrent in a single in-plane ferromagnet with spin-orbit
coupled contacts to superconductors},
  journal={Physical Review B},
  volume={100},
  pages={224519},
  year={2019},
  publisher={APS},
  doi = {10.1103/PhysRevB.100.224519}
}

@article{banerjee18,
  author={Banerjee, N. and Ouassou, J. A. and Zhu, Y. and Stelmashenko, N. A. and Linder, J. and Blamire, M. G.},
  title={Controlling the superconducting transition by spin-orbit coupling},
  journal={Physical Review B},
  volume={97},
  pages={184521},
  year={2018},
  publisher={APS},
  doi = {10.1103/PhysRevB.97.184521}
}

@article{bregazzi2024,
  title={Enhanced controllable triplet proximity effect in superconducting spin–orbit coupled spin valves with modified superconductor/ferromagnet interfaces},
  author={A. T. Bregazzi; J. A. Ouassou; A. G. T. Coveney ; N. A. Stelmashenko; A. Child; A. T. N'Diaye; J. W. A. Robinson ; F. K. Dejene ; J. Linder ; N. Banerjee},
  journal={Applied Physics Letters},
  volume={124},
  pages={162602},
  year={2024},
  publisher={AIP},
  doi = {10.1063/5.0209305}
}

@article{satchell18,
  author={Satchell, N. and Birge, N. O.},
  title={Supercurrent in ferromagnetic {J}osephson junctions with heavy metal interlayers},
  journal={Physical Review B},
  volume={97},
  pages={214509},
  year={2018},
  publisher={APS},
  doi = {10.1103/PhysRevB.97.214509}
}

@article{chiodi12,
  author={Chiodi, F. and Ferrier, M. and Gu\'{e}ron, S. and Cuevas, J. C. and Montambaux, G. and Fortuna, F. and Kasumov, A. and Bouchiat, H.},
  title={Geometry-related magnetic interference patterns in long SNS {J}osephson junctions},
  journal={Physical Review B},
  volume={86},
  pages={064510},
  year={2012},
  publisher={APS},
  doi = {10.1103/PhysRevB.86.064510}
}

\end{multicols}

\end{document}